\documentclass[preprint]{aastex701}
\hypersetup{linkcolor=blue,citecolor=blue,filecolor=blue,urlcolor=blue}
\usepackage{amsmath}
\usepackage{bm}

\begin{document}

\title{Spectral Microlensing of Extragalactic H\,II Regions by Stellar-Mass Black Holes}

\author[orcid=0000-0002-0409-5719,gname="Dezi",sname='Liu']{Dezi Liu}
\affiliation{South-Western Institute For Astronomy Research, Yunnan University,  Kunming 650500, China}
\affiliation{Key Laboratory of Survey Science of Yunnan Province, Yunnan University, Kunming 650500, China}
\email[show]{adzliu@ynu.edu.cn}

%\collaboration{all}{The Terra Mater collaboration}

\begin{abstract}
Most of the Milky Way's predicted stellar-mass black holes remain hidden, especially at high Galactic latitudes or in the Galactic halo, where traditional dense-field stellar microlensing is ineffective. We propose an alternative method to map this isolated population via the spectral microlensing of compact, extragalactic H\,II regions. Projected onto the source plane, the physical Einstein radius of a Galactic black hole can  match the typical core sizes of H\,II regions in distant galaxies. Microlensing triggers an achromatic magnification, producing distinct narrow emission-line excesses in integrated galaxy spectra. Because gravitational lensing is wavelength-independent, intrinsic line ratios are preserved, offering a robust discriminant against false-positive astrophysical transients. Notably, the efficiency of this method depends critically on the size of the H\,II regions: while extended regions suffer from low optical depth, compact regions with a physical size $\lesssim 10$ pc offer significantly higher magnifications. These compact cores, however, are heavily dust-obscured at optical wavelengths, making infrared and radio observations the primary windows for this method. Even so, the spatial sparseness of background H\,II regions and the stringent alignment requirement for high magnification limit the expected event rate to $\sim 10^{-6}$ per year. Nevertheless, this method offers a unique opportunity to detect stellar-mass black holes and constrain their abundance in such low-density environments.
\end{abstract}



\section{Introduction}
Stellar-mass black holes (BHs) with masses in the range $3$–$150\,M_\odot$ represent the natural endpoint of massive star evolution \citep{2012ApJ...749...91F, 2016ApJ...821...38S}. 
% A black hole forms when the collapsing core of a massive star exceeds the maximum mass of a neutron star. The formation efficiency of SMBHs depends sensitively on the progenitor's metallicity, rotation, and binary interactions \citep{2016A&A...594A..97B, 2020MNRAS.499.3214M}. 
Population synthesis models predict a total of $10^{8}$ to $10^{9}$ such BHs in the Milky Way, a large fraction of which are expected to be isolated \citep{1996ApJ...457..834T, 2020A&A...638A..94O}. Despite this theoretical abundance, fewer than 100 BHs have been identified in our Galaxy, mostly in X-ray binaries and astrometric binaries \citep{2025Symm...17.1393B}. The first unambiguous isolated BH was detected through a combination of photometric and astrometric microlensing \citep{2022ApJ...933...83S, 2025ApJ...983..104S}. This object has a mass of $7.15\,M_\odot$, lies at a distance of $1.52\,\mathrm{kpc}$, and moves with a space velocity of $51.1\,\mathrm{km\,s^{-1}}$ relative to local stars, indicating a modest natal kick. This detection confirms that isolated BHs exist in the Galactic disk, yet the vast majority remain hidden.

Several techniques have been used to search for BHs in the Milky Way. Astrometric microlensing monitors the apparent position shift of a background star when a compact object passes in front of it \citep{1995A&A...294..287H}. Combined with the photometric light curve, this method allows a direct determination of the lens mass and distance. With superb astrometric precision, space telescopes such as HST and Gaia have enabled several BH candidates to be identified \citep[e.g.,][]{2022ApJ...933...83S,2022ApJ...933L..23L,2022ApJ...937L..24M,2023MNRAS.518.1057E,2023MNRAS.521.4323E,2024A&A...686L...2G,2025ApJ...983..104S}. X-ray observations provide another powerful tool for detecting BHs in binary systems, where the accretion of material from a companion star releases gravitational potential energy, producing copious X-ray emission. To date, around seventy BHs have been identified \citep{2025Symm...17.1393B}. More recently, \citet{2025ApJ...988L..12M} proposed detecting isolated stellar-mass BHs 
via their electromagnetic emission. By generating synthetic model spectra of isolated BHs accreting from various interstellar medium environments, they showed that such emission may be detectable under the sensitivities of current facilities such as ALMA \citep{2009IEEEP..97.1463W} and JWST \citep{2006SSRv..123..485G}. 
A similar mechanism was also investigated by \citet{2020ApJ...901...39G}, who demonstrated that low-density accretion flows onto wandering BHs with masses of $\gtrsim 10^{5}\,M_\odot$ orbiting in the outskirts of their host galaxies can produce radiation spectra peaking at millimeter wavelengths.

%Understanding the population of stellar-mass black holes is of great astrophysical importance. For example, their mass function and spatial distribution directly probe supernova explosion mechanisms and natal kicks \citep{2012MNRAS.425.2799R, 2020MNRAS.499.3214M, 2024ApJ...963...63B}. Moreover, these black holes play crucial dynamical roles in dense stellar environments \citep{2025ApJ...991..146R, 2025A&A...695L..19H} and provide essential constraints on binary evolution models \citep{2021hgwa.bookE..16M, 2004ApJ...612.1044P}. These studies remain statistically limited due to the small sample of known SMBHs. Therefore, developing various detection methods to build larger SMBH samples is highly desirable.

Most existing BH searches rely on high background stellar densities (e.g., in the Galactic disk, bulge, or Magellanic Clouds), as only there does the detection probability become appreciable. This necessity, however, limits the accessible spatial distribution, particularly impeding the detection of BHs at high Galactic latitudes or in the halo. To address this limitation, in this Letter we propose an alternative approach based on spectral microlensing of extragalactic H\,II regions. Typically, a galaxy contains hundreds to thousands of such regions. Individual H\,II regions span a wide range of sizes in diameters, from compact cores ($\lesssim1$ pc) to more extended structures ($10$--$100$ pc) \citep{2021MNRAS.505.2801L, 2023MNRAS.520.4902G, 2025AA...696A..78B, 2026A&A...706A..95B}. H\,II regions are powerful emitters of strong emission lines. Compact H\,II regions are often heavily obscured by dust and thus are not visible at optical wavelengths, but they emit strongly in the infrared and radio bands \citep{2021MNRAS.505.2801L, 2023A&A...678A.129B}. In contrast, more extended H\,II regions are less attenuated and exhibit prominent emission lines at optical wavelengths, most notably H$\alpha$, [O\,III] $\lambda5007$, and H$\beta$, which dominate the integrated spectra of star-forming galaxies \citep{2025arXiv250217680I}.

Extensive studies have shown that the H$\alpha$ luminosity function of H\,II regions follows a power law with a slope of $\alpha \approx 1.5$--$2.0$ \citep{1989ApJ...337..761K,2022A&A...658A.188S,2025AA...696A..78B}. Rough estimates indicate that individual H\,II regions typically contribute about $0.1\%$ to $10\%$ of a galaxy's total H$\alpha$ emission, depending on the galaxy's star formation rate and stellar mass. In massive spiral galaxies, the brightest H\,II region accounts for $1\%$--$10\%$ of the total, whereas in dwarf systems, a single H\,II region can dominate the H$\alpha$ emission, contributing up to $50\%$--$100\%$ \citep{2025AA...696A..78B}.

Star-forming galaxies are numerous in the Universe and can be readily observed at high Galactic latitudes. We will demonstrate that their H\,II regions, particularly the compact ones, have sizes that can be comparable to or smaller than the Einstein radius projected at the source plane for typical BHs. Consequently, these H\,II regions can be strongly magnified, producing observable variations in their characteristic emission lines. 
%Because gravitational lensing is achromatic, all emission lines originating from the same H\,II region are magnified equally, preserving their intrinsic line ratios. By comparing spectra obtained at different epochs, the spectral difference reveals narrow flux excesses in each of these lines — a clean multi-line signature that unambiguously distinguishes lensing effect from other variable phenomena. 
This Letter is structured as follows. Section~\ref{sec:model} presents the spectral microlensing model and its effects on extragalactic H\,II regions. We then discuss detectability and the expected number of events based on current observations. Section~\ref{sec:cc} concludes the results and briefly comments on observing strategies. We adopt a standard $\Lambda$CDM cosmology with $H_0 = 72$ km s$^{-1}$ Mpc$^{-1}$, $\Omega_m = 0.3$, and $\Omega_\Lambda = 0.7$.

%Two-epoch spectroscopic data for hundreds of thousands of galaxies are already available from archival surveys such as the Sloan Digital Sky Survey (SDSS) at $z \sim 0.1$ and the Dark Energy Spectroscopic Instrument (DESI), which are well-suited to test this method \citep{York2000, DESI2016}. This approach works toward any extragalactic direction and offers a promising new pathway to unveil the hidden population of isolated black holes in our Galaxy.

\section{Spectral Microlensing}\label{sec:model}
In this section, we quantify the spectral microlensing effect of extragalactic H\,II regions based on standard gravitational lensing theory. A foreground BH can be treated as a point mass lens with mass $M$. Its Einstein radius is given by
\begin{equation}
\theta_E = \sqrt{\frac{4GM}{c^2} \frac{D_{LS}}{D_L D_S}},
\label{eq:einstein}
\end{equation}
where $D_L$, $D_S$, and $D_{LS}$ denote the angular diameter distances to the lens, to the source, and between the lens and the source, respectively. $G$ is the gravitational constant, and $c$ is the speed of light. Since the distance to extragalactic H\,II regions ($D_S \sim \mathrm{Mpc}$ to $\mathrm{Gpc}$) is much larger than the distance to the lensing BH ($D_L \sim \mathrm{kpc}$), we have $D_{LS} \approx D_S$. Equation (\ref{eq:einstein}) then reduces to
\begin{equation}
\begin{split}
\theta_E &\approx \sqrt{\frac{4GM}{c^2} \frac{1}{D_L}} \\ 
&= 9\,\left(\frac{M}{10.0M_\odot}\right)^{1/2}\left(\frac{D_{L}}{1.0\,\mathrm{kpc}}\right)^{-1/2}\,\mathrm{mas}.
\label{eq:einstein_simple}
\end{split}
\end{equation}
Thus, $\theta_E$ depends primarily on the lens mass and its distance. For typical BHs in the Milky Way, $\theta_E$ is expected to be on the order of milliarcseconds. As an illustration, using the measured parameters of the first confirmed isolated BH \citep[OGLE-2011-BLG-0462;][]{2025ApJ...983..104S}, 
%with mass $M = 7.15\,M_\odot$ and distance $D_L = 1.52$\,kpc \citep{2025ApJ...983..104S}, 
we estimate the Einstein radius to be $\theta_E \approx 6.19$\,mas.

To achieve significant magnification, the physical size $R_{S}$ of the background H\,II region should satisfy $R_{S} \lesssim \theta_E D_S$, or equivalently $\theta_S \lesssim \theta_E$, where $\theta_S = R_S/D_S$ is the angular size of the source. Assuming a uniform disk intensity distribution, the peak magnification is achieved when the source center is perfectly aligned with the lens \citep{1994ApJ...430..505W}:
\begin{equation}
\mu_{\text{peak}} = \sqrt{ 4\left(\frac{\theta_E}{\theta_{S}}\right)^2 + 1 }.
\label{eq:mmax}
\end{equation}
When $\theta_E = \theta_{S}$, the peak magnification is $\mu_{\text{peak}} = \sqrt{5} \approx 2.24$. This sets a minimum magnification threshold that can be expected under perfect alignment. We note that real H\,II regions possess complex internal structures,  such as clumps, filaments, and radial density gradients,  which can affect the exact magnification. Detailed modeling would require knowledge of the brightness profile of individual H\,II regions, which is generally unavailable for distant galaxies. Nevertheless, Equation (\ref{eq:mmax}) serves as a useful first-order approximation for estimating the typical magnification scale. 
%In our following analysis, we assume perfect alignment for simplicity, noting that any misalignment would result in a magnification smaller than $\mu_{\text{peak}}$.

\begin{figure}
    \centering
    \includegraphics[width=1.0\linewidth]{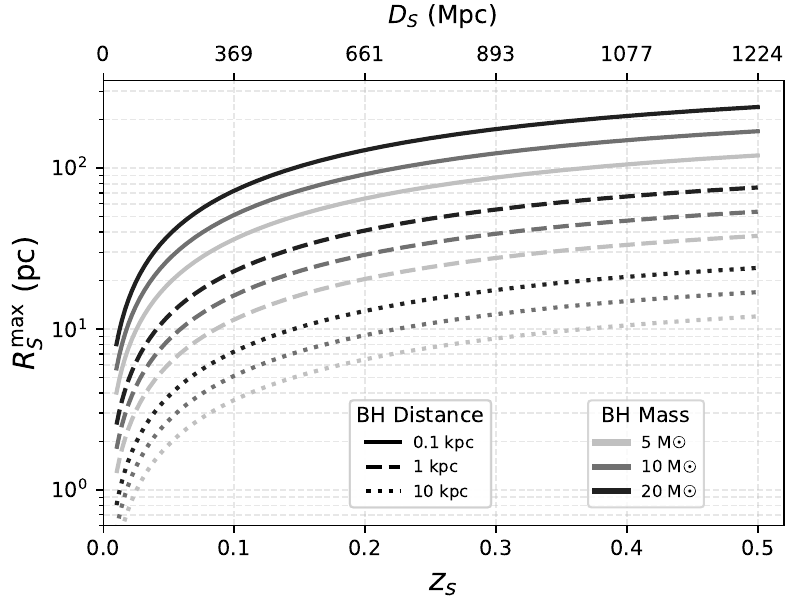}
    \caption{Maximum physical size of H\,II regions $R_{S}^{\mathrm{max}}$ versus source redshift $z_s$ for different BH masses (5, 10, 20~$M_{\odot}$) and distances (0.1, 1, 10~kpc), with the top x-axis showing the corresponding angular diameter distance $D_S$ in Mpc. $R_{S}^{\mathrm{max}}$ defines the upper size limit of the H\,II region within which significant lensing magnification can be produced.}
    \label{fig:HIIsize}
\end{figure}

Figure~\ref{fig:HIIsize} shows the maximum physical size ($R_{S}^{\mathrm{max}} \equiv \theta_E D_S$) of H\,II regions as a function of source redshift $z_s$. Because $\theta_E$ is approximately independent of $z_S$, the redshift dependence of $R_{S}^{\mathrm{max}}$ arises mainly from $D_S$. Across the entire redshift range considered ($0.01 \leq z_S \leq 0.5$), $R_{S}^{\mathrm{max}}$ remains well within the characteristic size scales of observed H\,II regions, indicating that such regions are readily magnified under typical lensing conditions. 
%In general, massive and nearby SMBHs present larger magnification for a broader range of H\,II region size scales, from compact to extended.

\begin{figure}
    \centering
    \includegraphics[width=1.0\linewidth]{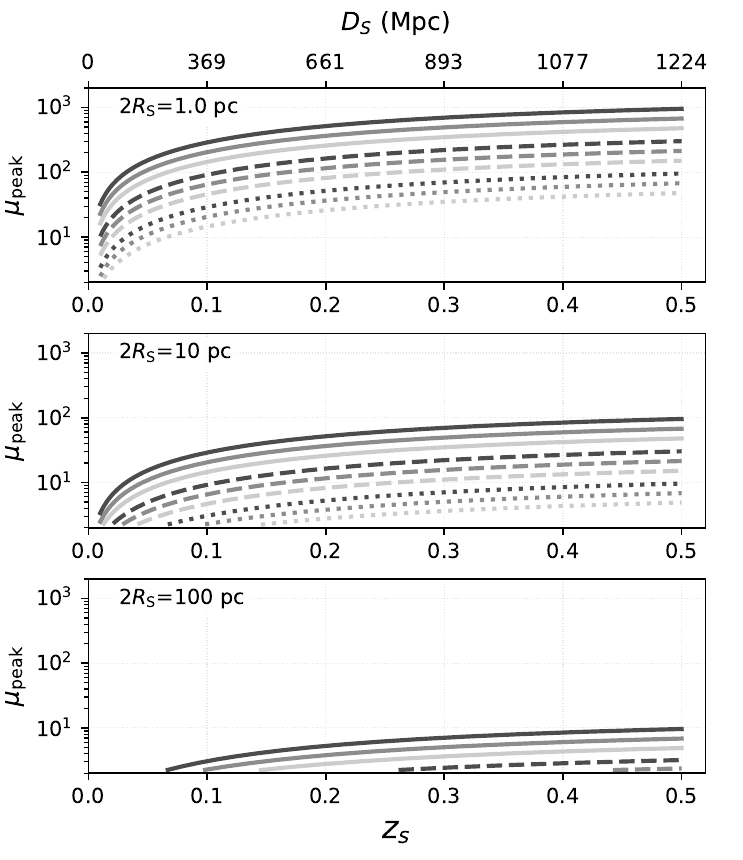}
    \caption{Peak magnification $\mu_{\text{peak}}$ as a function of source redshift $z_S$ for different H\,II region sizes (diameters of $2R_{S}$=1, 10, and 100 pc), assuming perfect alignment between the source and the lens. Different line styles and grayscales represent various BH masses and distances, following the same convention as in Figure~\ref{fig:HIIsize}. Compact H\,II regions (1\,pc) exhibit magnifications $\gtrsim 10$ for nearly all lens configurations, while more extended H\,II regions (10--100\,pc) achieve comparable magnifications only for the most massive and nearby BHs.}
    \label{fig:mupeak}
\end{figure}

Figure~\ref{fig:mupeak} shows the peak magnification as a function of redshift for fixed H\,II region sizes. Three representative diameters ($2R_S$) are adopted: 1\,pc, 10\,pc, and 100\,pc. For compact H\,II regions ($2R_S = 1\,$pc), the magnification exceeds a factor of 10 for nearly all combinations of BH mass and distance under consideration. It should be noted, however, that such compact sources are often severely attenuated by dust, rendering them difficult to detect in optical observations. Instead, they are more observable at infrared or radio wavelengths. For larger H\,II regions (10--100\,pc), the massive and nearby BHs can still achieve magnifications of order 10. These larger systems suffer less from dust extinction and are therefore promising targets for detection via optical spectroscopy.

\subsection{Effect on Emission Lines from an H\,II Region}
H\,II regions produce both continuum and strong emission line radiation. In galaxies, the continuum is primarily dominated by large-scale stellar populations and diffuse ionized gas, with spatial extents far larger than the Einstein radius of the lensing BH projected at the source, while the fractional contribution of continuum from H\,II regions is minimal \citep{2025arXiv250217680I}. In contrast, emission lines from individual H\,II regions can contribute up to 0.1\%--10\% of the total line emission, depending on the galaxy type and H\,II region luminosity. Consequently, for the integrated galaxy spectra, the magnification effect is expected to be significant for emission lines but negligible for the continuum. We therefore focus exclusively on the lensing magnification of emission lines from H\,II regions.

Gravitational lensing is achromatic: the magnification $\mu$ is independent of wavelength. We assume that all emission lines from the same H\,II region are magnified by the same factor $\mu$. Let $F_{i}$ be the total unlensed intensity of the entire galaxy in line $i$ (e.g., H$\alpha$, [OIII], H$\beta$), and let the unlensed intensity of a single H\,II region in that line be a fraction $\epsilon_i$ of the total, i.e., $F_{\text{H\,II},i} = \epsilon_i F_{i}$ with $0 < \epsilon_i \ll 1$. The intensity from the rest of the galaxy (other H\,II regions, diffuse ionized gas) is then $F_{\text{other},i} = (1 - \epsilon_i) F_{i}$.

After lensing, only the source H\,II region is magnified, while the rest of the galaxy remains unchanged. The lensed total intensity is therefore:
\begin{equation}
F_{i}^{\text{lensed}} = F_{\text{other},i} + \mu F_{\text{H\,II},i}.
\end{equation}
Subtracting the unlensed total intensity $F_{i} = F_{\text{other},i} + F_{\text{H\,II},i}$ from the lensed intensity yields the intensity change. We define the relative change as:
\begin{equation}
m_{i} \equiv \frac{F_{i}^{\text{lensed}} - F_{i}}{F_{i}} = \frac{\Delta{F_{i}}}{F_{i}} = (\mu - 1) \epsilon_i.
\label{eq:line_magnification}
\end{equation}
Thus, $m_{i}$ represents the fractional magnification excess of the total galaxy intensity in line $i$ due to spectral microlensing.

A powerful diagnostic follows from the achromaticity of lensing: the same factor $(\mu-1)$ applies to all lines. For any two lines $i$ and $j$, the ratio of the relative changes is:
\begin{equation}
\frac{m_i}{m_j} = \frac{\epsilon_i}{\epsilon_j}.
\label{eq:line_ratios}
\end{equation}
That is, the ratio of the observed magnification excess is independent of $\mu$ and depends only on the intrinsic line intensity ratios of the H\,II region relative to the total galaxy emission. This property provides a robust discriminant against false positives, as different astrophysical processes (e.g., supernovae, active galactic nuclei, and variable stars) or calibration artifacts would produce intensity ratios that vary between different lines and epochs. Multi-line consistency, therefore, can serve as a standard for identifying lensing events caused by BHs. Note that, in reality, different emission lines may originate from slightly different spatial zones within an H\,II region due to its ionization structure, and spatial variations in dust attenuation can effectively modify their relative distributions \citep[e.g.,][]{2013ApJ...778L..41L, 2016MNRAS.462.1757G}. In principle, when the lens caustic resolves internal sub-structures in the high-magnification regime ($\theta_{\mathrm{S}} \lesssim \theta_{\mathrm{E}}$), this spatial segregation could induce mild differential magnification between different emission lines. However, we expect this effect to be secondary. First, line-of-sight projection in integrated two-dimensional observations causes the emission zones to substantially overlap, reducing effective spatial offsets. Second, even if minor differential magnification occurs, it would manifest as a modest deviation in line ratios rather than mimic an astrophysical transient, although a precise quantitative bound on this effect would require detailed photoionization and radiative transfer modeling.

Furthermore, as discussed above, the magnification induced by BHs is primarily imprinted on emission lines, while the continuum remains largely unaffected. This behavior is markedly different from other known types of transients, which typically exhibit strong broad-band photometric flux variations. For the spectral microlensing events considered here, broad-band photometric surveys are largely insensitive to the signal. Instead, spectroscopic observations or narrow-band imaging that target emission lines provide the necessary sensitivity.

Observationally, applying the multi-line diagnostic requires multi-epoch data. In a single epoch, the fractional contribution $\epsilon_i$ is generally unknown. However, when comparing a lensed epoch to an unlensed baseline epoch, the measured fractional magnification excess $m_i$ can be directly determined. This allows one to identify a promising candidate by detecting coherent changes in multiple emission lines while the continuum remains unchanged. Once a candidate is identified, follow-up observations can monitor the microlensing light curve, during which all emission lines should trace the same temporal magnification pattern. The ratios of the excess fractions (Equation~\ref{eq:line_ratios}) provide a powerful confirmation, as they should remain constant across different lines and epochs. This two-stage strategy, i.e., candidate selection via sparse observations followed by targeted monitoring, offers a practical path to search for microlensing events.

\subsection{Detectability Analysis}\label{sec:da}
We now quantify the observational prospects for detecting spectral microlensing events from BHs.

Considering only shot noise, the detection significance $S_i$ (defined as $\Delta F_i / \sigma_{\Delta F_i}$) satisfies $(\mu-1)\epsilon_i \approx \sqrt{2} S_i / \mathrm{SNR}_i$, where $\mathrm{SNR}_i$ is the signal-to-noise ratio of the unlensed line intensity. Assuming a moderate fractional contribution of $\epsilon_i \sim 1\%$, and taking $\mu \sim 10$, the relative change in the line intensity is approximately $9\%$ according to Equation~(\ref{eq:line_magnification}). 
A detection significance exceeding $7\sigma$ would then require $\mathrm{SNR}_{i} \gtrsim 110$.
Thus, these intensity variations should be detectable in spectra with sufficiently high SNRs. For events with lower magnification, however, the excess becomes considerably smaller and may be lost in the noise, making detection challenging. In this context, the most promising targets are face-on star-forming galaxies, where the number of H\,II regions is larger and blending effects are minimized.

A caveat concerns the time-dependent nature of magnification. In practice, the magnification varies as the foreground BH moves relative to the background H\,II region. For most of the event duration, the magnification remains below its peak value. Moreover, perfect alignment between the lens and the source is rarely achieved, so the peak magnification is seldom attained in real observations. Both effects reduce the effective detectability of lensing events. An additional consideration is blending. Higher source redshift does not necessarily yield stronger magnification, because the increase in angular diameter distance leads to enhanced blending of the background H\,II region with neighboring H\,II regions or extended continuum-emitting structures. Such blending effectively enlarges the observed source size and dilutes the lensing signal, thereby suppressing the magnification. Even for H\,II regions at moderate redshifts, line-of-sight structures within the source galaxy can also spatially dilute the light distribution, similarly reducing the achievable magnification.

We proceed to estimate the expected number of detectable events. The optical depth $\tau$ is defined as the probability that a given background source lies within the Einstein radius of a foreground lens at any instant \citep[e.g.,][]{2008arXiv0811.0441M}. For extragalactic H\,II regions lensed by Galactic BHs, the lenses are confined to the Milky Way. Different regions of the Milky Way exhibit different BH densities \citep{2020A&A...638A..94O}. Since observable H\,II regions are typically located in low-stellar-density environments or at high Galactic latitudes, we restrict our analysis to the contributions from the Galactic disk and halo along the line of sight. Let $L_{\text{disk}}$ denote the path length from the Sun (observer) to the disk edge, and assume a constant BH mass density $\rho_{\text{disk}}$ within the disk. Beyond the disk lies the halo, with a path length $L_{\text{halo}}$ and a constant mass density $\rho_{\text{halo}}$. 
Under the approximation $D_{LS} \approx D_S$, the optical depth is given by (Appendix~\ref{app:od})
\begin{equation}
\tau = \tau_\text{disk} + \tau_\text{halo},
\label{eq:tau}
\end{equation}
where 
\begin{align}
\tau_{\rm disk} &= \frac{2\pi G}{c^2} \rho_{\rm disk} L_{\rm disk}^2, \nonumber \\
\tau_{\rm halo}  &= \frac{2\pi G}{c^2} \rho_{\rm halo} \left( L_{\rm halo}^2 + 2 L_{\rm disk} L_{\rm halo} \right), \nonumber
\end{align}
represent the contributions from the Galactic disk and halo, respectively.
Based on the simulations of \citet{2020A&A...638A..94O}, the disk has a thickness of $0.3$ kpc and extends radially from $2$ to $15$ kpc, while the halo is modeled as a spherical shell spanning $15$ to $30$ kpc. A gap exists between the disk and the halo, which we include as part of the halo. The average BH mass density in the Galactic disk is approximately $\rho_{\text{disk}} = 4.55 \times 10^{-22} \, \text{kg} \, \text{m}^{-3}$, and that in the halo is $\rho_{\text{halo}} = 5.47 \times 10^{-26} \, \text{kg} \, \text{m}^{-3}$. For a line of sight perpendicular to the Galactic plane, we obtain $\tau \sim 2.49 \times 10^{-10}$. This is the probability that a given H\,II region is aligned within one Einstein radius of a foreground BH at any instant.

\begin{figure}
    \centering
    \includegraphics[width=1.0\linewidth]{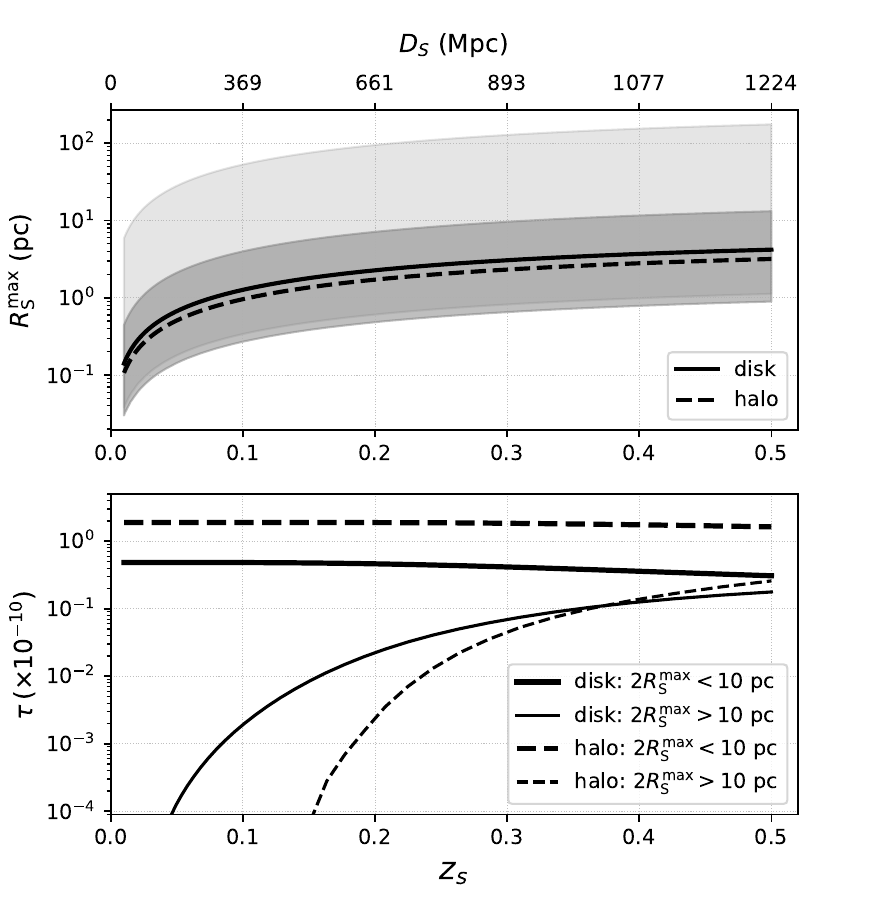}
    \caption{\textit{Top panel}: physical size upper limit $R_{S}^{\max}$ of background H\,II regions as a function of source redshift $z_{S}$ (or $D_{S}$). The solid and dashed black curves denote the mean $R_{S}^{\max}$ for BHs in the Galactic disk and halo, respectively. The light and dark shaded regions enclose the full range (from minimum to maximum) of $R_{S}^{\max}$ for each component. \textit{Bottom panel}: optical depth $\tau$ contributed by disk (solid lines) and halo (dashed lines) BHs. Thick lines are for compact H\,II regions, while thin lines are for extended ones.}
    \label{fig:tau_estimate}
\end{figure}

However, as shown in Figure~\ref{fig:mupeak}, when the BH is distant or the H\,II region is large, the peak magnification becomes small, making it difficult to produce a detectable microlensing signal in observations. Here we further estimate the optical depth under the condition $\mu_\text{peak}>10$. From Equation~(\ref{eq:mmax}), for a given BH mass and distance, we can determine the physical size upper limit $R_{S}^{\max}$ of an H\,II region at a given source distance $D_{S}$ such that actual H\,II regions with $R_{S}<R_{S}^{\max}$ will always have $\mu_\text{peak}>10$. The top panel of Figure~\ref{fig:tau_estimate} shows the results for the disk and halo calculated from the synthetic catalogs\footnote{\url{https://bhc.syntheticuniverse.org}} of \citet{2020A&A...638A..94O}. These catalogs yield distinct BH mass distributions and number counts: in the disk, the BH population amounts to $\sim10^{8}$ with an average mass of $\sim14\,M_{\odot}$; in the halo, the number is only $\sim3.5\times10^{6}$ but the average mass is larger, $\sim21\,M_{\odot}$. The black solid and dashed lines represent the mean values of $R_{S}^{\max}$, while the light and dark shaded regions indicate the ranges from the minimum to maximum $R_{S}^{\max}$ in the two respective zones. As seen from the figure, BHs in the disk can produce significant magnification for H\,II regions over a wide range of sizes ($\sim0.04$--$170\,\mathrm{pc}$), whereas those in the halo mainly affect compact H\,II regions ($\lesssim15\,\mathrm{pc}$).

We therefore classify H\,II regions by physical diameter into compact ($2R_{S}^{\max}<10\,\mathrm{pc}$) and extended ($2R_{S}^{\max}>10\,\mathrm{pc}$), and estimate the optical depth contributions from different Galactic regions by rejecting BHs that cannot produce $\mu_\text{peak}>10$. The bottom panel of Figure~\ref{fig:tau_estimate} displays the optical depth contributions for H\,II regions of different sizes. The results show that, for both the disk and the halo, the optical depths toward compact H\,II regions are comparable: $\tau_{\mathrm{disk}}\sim 0.5\times10^{-10}$ (thick solid line) and $\tau_{\mathrm{halo}}\sim 1.9\times10^{-10}$ (thick dashed line). For extended H\,II regions, however, the optical depths are lower by 1--4 orders of magnitude (thin solid and dashed lines). This indicates that compact H\,II regions are more efficient tracers for detecting Galactic BHs, although their observation requires infrared and radio facilities. For extended H\,II regions, one needs to target more distant galaxies ($z\gtrsim0.3$) where their angular sizes are smaller, yielding optical depths that are comparable to those of compact ones.

%where $n_{\text{BH}}$ is the number density of black holes in the Milky Way, and $L_{\text{MW}}$ is the characteristic scale over which lenses are distributed. Assuming a spatially uniform number density $n_{\text{BH}} = 1.0\times10^{-5}$ pc$^{-3}$ \citep{2020A&A...638A..94O,2025ApJ...988L..12M} and adopting $M = 10\,M_\odot$ and $L_{\text{MW}} = 15$ kpc, we obtain $\tau \sim 6.8\times10^{-9}$. This is the probability that a given H\,II region is aligned within one Einstein radius of a foreground SMBH at any instant.

The characteristic duration of a microlensing event is the Einstein crossing time:
\begin{equation}
\begin{split}
t_E &= \frac{\theta_E D_L}{v_{\perp}} \\
&= 78\,\left(\frac{v_{\perp}}{200 \,\text{km/s}}\right)^{-1}\left(\frac{M}{10\,M_\odot}\right)^{1/2}\left(\frac{D_{L}}{1\,\mathrm{kpc}}\right)^{1/2} \text{ days},
\label{eq:crossing_time}
\end{split}
\end{equation}
where $v_{\perp}$ denotes the relative transverse velocity. For typical parameters, the resulting timescale is on the order of a few months. 
The numerical estimate of $v_{\perp}$ is detailed in Appendix~\ref{app:tv}.

The event rate is derived by following the same equations in \citet{2008arXiv0811.0441M}, assuming that all Einstein crossing times are identical. Let $N_{\text{H\,II}}$ be the number of H\,II regions per galaxy. The expected number of detections across $N_{\text{gal}}$ galaxies is then:
\begin{align}
N_{\text{det}} &\approx \frac{2}{\pi} N_{\text{gal}} \, N_{\text{H\,II}} \, \frac{\tau}{t_E} \nonumber \\
&= 2.3\times 10^{-5} \, \left(\frac{N_{\text{gal}}}{1000}\right) \left(\frac{N_{\text{H\,II}}}{100}\right) \left(\frac{\tau}{10^{-10}}\right)  \left(\frac{t_{E}}{100\ \text{days}}\right)^{-1} \text{yr}^{-1}.
\label{eq:ndet}
\end{align}
Rough numerical estimates of $N_{\rm gal}$ and $N_{\rm H\,II}$ are detailed in Appendix~\ref{app:num}. Here, we have adopted a global value of the optical depth $\tau$. As discussed above, for extended H\,II regions, the actual event rate $N_{\rm det}$ is substantially lower than this estimate.
Unlike stars, galaxies are generally more sparsely distributed. Face-on star-forming galaxies, which typically contain a larger number of H\,II regions, serve as promising observational targets. Based on current observations of H\,II regions \citep[e.g.,][]{2023MNRAS.520.4902G, 2025AA...696A..78B,2026A&A...706A..95B},the predicted event rate is at most $10^{-5}$  per year, making real detection extremely challenging.

Furthermore, we note that our estimate of $N_{\rm det}$ implicitly assumes that any alignment within the full Einstein radius yields a detectable signal. However, achieving a high magnification requires the source to pass much closer to the lens. Let $\beta$ denote the angular separation between the lens and the source center, and define the normalized impact-parameter $b \equiv \beta/\theta_E$. Physically, $b<1$ means the source falls within the Einstein radius. The effective event rate is then reduced by a factor of $b$, so $N_{\rm eff, det} = b\, N_{\rm det}$ (see Appendix~\ref{app:ip} for the detailed derivation). For a threshold of $b = 0.1$ (roughly corresponding to $\mu = 10$), Equation~(\ref{eq:ndet}) gives a revised event rate of $N_{\rm eff, det} \sim 10^{-6}$ per year.
Despite this extremely low rate, the method still offers unique potential.
First, it targets BHs at high Galactic latitudes or in the Galactic halo, a regime largely inaccessible to existing techniques that achieve high optical depth only toward dense stellar fields. Spectral microlensing thus serves as a valuable complement. Second, even a null result would place independent upper limits on the abundance of isolated BHs in the environment, thereby constraining their diverse formation channels.

Two observational strategies can be pursued to search for these events. The first is a spectroscopic time-domain survey. While current photometric surveys such as LSST already enable multi-epoch observations over large sky areas across a decade \citep{LSST2019}, no spectroscopic time-domain survey has yet achieved this baseline. Looking ahead, advances in time-domain spectroscopy may enable such capabilities, thereby enhancing the method's practicality. The second strategy involves narrow-band imaging surveys (e.g., H$\alpha$), which offer higher efficiency and wider coverage, providing a more feasible path toward improving the event rate. A combination of narrow-band and broad-band observations can be employed: narrow-band imaging monitors variations in emission-line intensities, while broad-band imaging tracks changes in the continuum as a reference, since microlensing is not expected to cause continuum variations.

\section{Conclusion and comments}\label{sec:cc}
We have presented an alternative method to detect isolated BHs in the Milky Way via spectral microlensing of extragalactic H\,II regions. When a foreground BH aligns with a background H\,II region in a distant star-forming galaxy, the magnification produces narrow emission-line excesses while preserving intrinsic line ratios due to achromaticity. We demonstrate that the Einstein radius of BHs is comparable to typical H\,II regions, enabling detectable magnifications. However, our analysis reveals that practical detection remains extremely challenging due to the sparse distribution of available galaxies.

The era of large-area, multi-epoch sky surveys has arrived. Image subtraction techniques are routinely used in photometric surveys to discover transients. Similarly, such techniques can be developed for future time-domain spectroscopic surveys. Unlike image subtraction, which measures integrated flux changes, spectral subtraction provides wavelength-resolved information, enabling the detection of transients that may be completely obscured in photometric observations. This is particularly advantageous for events where flux changes are confined to specific spectral lines, such as the spectral microlensing investigated here. As an example, existing surveys like SDSS \citep[e.g.,][]{2020ApJS..249....3A,2022ApJS..259...35A} and DESI \citep{2016arXiv161100036D} have observed nearly ten thousand square degrees of the sky, with a substantial fraction of galaxies having multi-epoch spectra from both surveys, separated by roughly a decade.
While this extended baseline is too sparse to monitor continuous microlensing light curves, it provides an excellent multi-epoch dataset to initiate a search for discrete emission-line flux variations between widely separated eras.

Future multi-wavelength surveys will provide the complementary coverage needed to address the wide range of physical scales spanned by background H\,II regions. Specifically, the near-infrared capabilities of the Nancy Grace Roman Space Telescope \citep{2019arXiv190205569A} and next-generation radio interferometers such as the Square Kilometre Array \citep[SKA;][]{2009IEEEP..97.1482D} will probe compact, dust-obscured H\,II regions, which yield the highest magnifications. Conversely, the wide-field optical coverage of the Chinese Space Station Survey Telescope \citep[CSST;][]{2026SCPMA..6939501C} will target more extended, less attenuated systems. By spatially resolving these diverse populations and mitigating host-galaxy blending, these missions will enhance the observable line excess, thereby lowering the detection threshold and offering a unique opportunity to probe isolated stellar-mass BHs at high Galactic latitudes and in the Galactic halo.

%% Please use the acknowledgment and contribution environments. This will 
%% be anonomyized when the "anonymous" style option is used. 
\begin{acknowledgments}
Dezi Liu gratefully acknowledges the anonymous referee for constructive comments that improved the quality of this manuscript. This work was supported by the Young Talent Special Project of the Xingdian Yingcai Support Program of Yunnan Province and by the National Natural Science Foundation of China (NSFC) under grant No. 12103043.
\end{acknowledgments}

\appendix
\section{Optical depth}\label{app:od}
The optical depth is the probability that a given background source lies within the Einstein radius of any foreground lens at any instant. It is defined as
\begin{equation}
    \tau = \int_{0}^{D_{S}} n(D_{L})\, (\pi R_{E})^2 \, \text{d}D_{L},
\end{equation}
where $n(D_{L})$ is the number density of BHs along the line of sight, and $R_{E} = D_{L}\theta_{E}$ is the physical Einstein radius in the lens plane. With $D_{LS} \approx D_S$, the optical depth simplifies to
\begin{equation}
    \tau = \frac{4\pi G}{c^2} \int_{0}^{D_{S}} \rho(D_{L})\, D_{L} \, \text{d}D_{L}.
\end{equation}
where $\rho(D_{L})=n(D_L)M$ is the average matter density of BHs at a distance $D_{L}$ from the observer. 

The line of sight from the Sun to a background H\,II region first crosses the Galactic disk and then the halo, beyond which BHs are negligible. Let $L_{\text{disk}}$ and $L_{\text{halo}}$ be the path lengths through the disk and halo, and $\rho_{\text{disk}}$ and $\rho_{\text{halo}}$ their respective constant matter densities. Substituting these into the integral yields
\begin{equation}
\begin{split}
    \tau &= \frac{4\pi G}{c^2} \left[\rho_{\text{disk}} \int_{0}^{L_{\text{disk}}} D_{L} \, \text{d}D_{L} + \rho_{\text{halo}} \int_{L_{\text{disk}}}^{L_{\text{disk}}+L_{\text{halo}}} D_{L} \, \text{d}D_{L} \right] \\
    &= \frac{2\pi G}{c^2} \left[ \rho_{\text{disk}} L_{\text{disk}}^2 + \rho_{\text{halo}} \left( L_{\text{halo}}^2 + 2 L_{\text{disk}} L_{\text{halo}} \right) \right].
\end{split}    
\end{equation}
For a line of sight perpendicular to the Galactic plane, the simulations of \citet{2020A&A...638A..94O} give a vertical BH distribution of $z \pm 0.15$ kpc in the disk. Assuming the Sun lies in the Galactic plane, we have $L_{\text{disk}} = 0.15$ kpc. The same simulations define the halo as a spherical shell from 15 to 30 kpc, with a gap between the disk and halo that we include as part of the halo. With the Sun at $L_{\odot} = 8$ kpc from the Galactic center, the line-of-sight distance to the outer halo boundary is $L_{\text{los}} = \sqrt{L_{\text{outer}}^{2} - L_{\odot}^2} \approx 28.91$ kpc, giving $L_{\text{halo}} = L_{\text{los}} - L_{\text{disk}} \approx 28.76$ kpc.

\section{Transverse velocity}\label{app:tv}
We derive the transverse velocity \(v_{\perp}\) of a BH as seen from the Sun, given its Galactocentric position vector \(\bm{R}_{\rm BH}\) and velocity vector \(\bm{V}_{\rm BH}\) in three spatial dimensions. The Sun's position and velocity in the same frame are \(\bm{R}_\odot\) and \(\bm{V}_\odot\). In this work, we adopt \(\bm{R}_\odot = (8.0, 0.0, 0.0)\ \text{kpc}\) and \(\bm{V}_\odot = (0.0, 220.0, 0.0)\ \text{km s}^{-1}\). The distance vector from the Sun to the BH is \(\bm{D} = \bm{R}_{\rm BH} - \bm{R}_\odot\), and the scalar distance is \(D = |\bm{D}|\). The unit vector along the line of sight is defined as \(\hat{\bm{r}} = \bm{D} / D\).

The velocity of the BH relative to the Sun is
\begin{equation}
    \bm{V}_{\rm rel} = \bm{V}_{\rm BH} - \bm{V}_\odot .
\end{equation}
We decompose \(\bm{V}_{\rm rel}\) into components parallel and perpendicular to \(\hat{\bm{r}}\). The parallel (radial) and perpendicular (transverse) components are
\begin{equation}
    \bm{V}_{\parallel} = (\bm{V}_{\rm rel} \cdot \hat{\bm{r}}) \, \hat{\bm{r}}, \qquad
    \bm{V}_{\perp} = \bm{V}_{\rm rel} - \bm{V}_{\parallel}
                  = \bm{V}_{\rm rel} - (\bm{V}_{\rm rel} \cdot \hat{\bm{r}}) \, \hat{\bm{r}} .
\end{equation}
This decomposition is unique and satisfies \(\bm{V}_{\perp} \cdot \hat{\bm{r}} = 0\), where \(\cdot\) denotes the dot product.

The transverse velocity is the magnitude of the perpendicular component:
\begin{equation}
    v_{\perp} = |\bm{V}_{\perp}| .
\end{equation}
Using the orthogonality \(\bm{V}_{\parallel} \perp \bm{V}_{\perp}\), we have \(|\bm{V}_{\rm rel}|^2 = |\bm{V}_{\parallel}|^2 + |\bm{V}_{\perp}|^2\). Thus,
\begin{equation}
    v_{\perp}
    = \sqrt{|\bm{V}_{\rm rel}|^2 - |\bm{V}_{\parallel}|^2}
    = \sqrt{|\bm{V}_{\rm rel}|^2 - (\bm{V}_{\rm rel} \cdot \hat{\bm{r}})^2} .
\end{equation}
Substituting the definitions of \(\bm{V}_{\rm rel}\) and \(\hat{\bm{r}}\), the transverse velocity is
\begin{equation}
    v_{\perp}
    =
    \sqrt{
        \left|\bm{V}_{\rm BH} - \bm{V}_\odot\right|^2
        -
        \left[
            \left(\bm{V}_{\rm BH} - \bm{V}_\odot\right)
            \cdot
            \frac{\bm{R}_{\rm BH} - \bm{R}_\odot}
                 {|\bm{R}_{\rm BH} - \bm{R}_\odot|}
        \right]^2
    }.
\end{equation}

The simulated catalogs generated by \citet{2020A&A...638A..94O} provide the Galactocentric position  \(\bm{R}_{\rm BH}\) and velocity  \(\bm{V}_{\rm BH}\) for BHs in the Galactic disk and halo. \(\bm{V}_{\rm BH}\) is the sum of the rotational motion in the Galactic potential and the natal kick velocity imparted by physical processes such as supernova explosions. Based on these data and the transverse velocity equation derived above, we obtain the following mean transverse velocities for single BHs: \(\sim 206\ \mathrm{km\,s^{-1}}\) for the Galactic disk and \(\sim 255\ \mathrm{km\,s^{-1}}\) for the Galactic halo. Therefore, we conservatively adopt \(v_{\perp} = 200\ \mathrm{km\,s^{-1}}\) to estimate the Einstein cross time $t_{E}$.

\section{Population estimate of H II regions}\label{app:num}

\begin{table*}
\centering
\caption{Summary of major H\,II region surveys and catalogs.}
\label{tab:hii_surveys}
\begin{tabular}{llcccc}
\hline
\hline
Survey & Telescope/Instrument & \(N_\mathrm{gal}\) & \(N_\mathrm{H\,II}\) & Redshift range & H\(\alpha\) Sensitivity\tablenotemark{a} \\
\hline
AMUSING++\tablenotemark{b}  & VLT/MUSE & 539  &  52,371 & \(0.004 < z < 0.06\) & \(\sim 10^{-17}\) \\
CALIFA\tablenotemark{c}      & Calar Alto 3.5m/PMAS & 924  & 26,408          & \(z \lesssim 0.08\)  & \(\sim 10^{-17}\) \\
PHANGS-MUSE\tablenotemark{d} & VLT/MUSE & 19   & 23,244          & \(z \lesssim 0.005\) & \(\sim 1 \times 10^{-18}\) \\
VESTIGE\tablenotemark{e}     & CFHT/MegaCam & 322  & 76,645          & \(z \sim 0.004\)     & \(\sim 2 \times 10^{-18}\) \\
\hline
\hline
\end{tabular}
\tablenotetext{a}{Typical \(3\sigma\) surface brightness detection limit in units of \(\mathrm{erg~s^{-1}~cm^{-2}~arcsec^{-2}}\).}
\tablerefs{
\(^{b}\)~\citet{2024MNRAS.528.6099L}; 
\(^{c}\)~\citet{2020MNRAS.494.1622E}; 
\(^{d}\)~\citet{2023MNRAS.520.4902G}; 
\(^{e}\)~\citet{2025AA...696A..78B,2026arXiv260502648B}.}
\end{table*}

\citet{2026A&A...706A..95B} utilized high-resolution HST narrow-band H$\alpha$ imaging to observe 19 nearby star-forming galaxies ($D < 20$~Mpc; $z \lesssim 0.005$; see Table~\ref{tab:hii_surveys}) from the PHANGS--MUSE survey \citep{2022A&A...659A.191E}, achieving a physical resolution of a few parsecs to 10~pc, an order of magnitude improvement over the ground-based MUSE observations. This combined dataset yields a catalog of approximately 5,000 spatially resolved H\,II regions with both photometric and spectroscopic measurements. We use this catalog to estimate $N_{\mathrm{H\,II}}$, where for each H\,II region, the PSF-deconvolved moment-based radius ($r_{\mathrm{mom,deconv}}$) is adopted as a measure of its intrinsic physical size. To ensure a consistent comparison between the two datasets, we apply identical selection criteria: 2$r_{\mathrm{mom,deconv}} < 100$~pc and $\mathrm{SNR}_{\mathrm{H}\alpha} > 100$, where $\mathrm{SNR}_{\mathrm{H}\alpha}$ is defined as the ratio of the H$\alpha$ flux to its corresponding uncertainty, computed independently from the HST and MUSE observations. At the typical distance of this nearby sample ($D \approx 15$~Mpc), this $\mathrm{SNR}_{\mathrm{H}\alpha}$ cut corresponds to a physical H$\alpha$ line-flux limit of $F_{\mathrm{H}\alpha} \gtrsim 10^{-15}~\mathrm{erg~s^{-1}~cm^{-2}}$ (equivalent to an intrinsic H$\alpha$ luminosity limit of $L_{\mathrm{H}\alpha} \gtrsim 10^{37}~\mathrm{erg~s^{-1}}$), isolating the brightest and most robust H\,II regions. Applying these cuts yields 1,949 H\,II regions from the HST data and 3,148 from the MUSE data, corresponding to average values of $N_{\mathrm{H\,II}} \sim 103$ and $\sim 166$ per galaxy, respectively. It is worth noting that these estimates are based on nearby galaxies. For more distant galaxies, achieving such high SNR will be more challenging, which would lead to a decrease in $N_{\mathrm{H\,II}}$.

To estimate the number of available galaxies \(N_\mathrm{gal}\) for H\,II region studies, we compiled the galaxy counts from several major H\,II surveys and catalogs in the literature (Table~\ref{tab:hii_surveys}). Despite some overlap among these surveys, the total number of unique galaxies is expected to exceed 1000. These galaxies are relatively nearby (\(D \lesssim 300\,\mathrm{Mpc}\); $z \lesssim 0.08$) and are primarily observed with ground-based integral field spectrographs or deep narrow-band imagers, where the spatial resolution is typically limited to \(\sim1\arcsec\) \citep{2020MNRAS.494.1622E,2024MNRAS.528.6099L}.
For observations with higher angular resolution, such as those from current and upcoming space-based telescopes including HST, JWST, and Roman, which can achieve resolutions of \(0.1\arcsec\) or better, H\,II regions can be resolved in galaxies at larger distances, significantly expanding the pool of potential targets. 

Nevertheless, for the purpose of our numerical estimates, we conservatively adopt \(N_{\mathrm{H\,II}} = 100\) and \(N_\mathrm{gal} = 1000\) for the detection number \(N_{\mathrm{det}}\) in Equation~(\ref{eq:ndet}).

\section{DERIVATION OF THE IMPACT-PARAMETER CORRECTION}\label{app:ip}

We derive here the impact-parameter correction used in estimating the event rate. Let $\beta$ denote the angular separation between the lens and the center of the source (H\,II region), and define the normalized impact parameter as
\begin{equation}
b \equiv \frac{\beta}{\theta_E},
\end{equation}
where $\theta_E$ is the Einstein radius. The normalized source radius is $p \equiv \theta_S/\theta_E$, as used in Equation~(\ref{eq:mmax}).

For a uniformly bright disk source, the magnification at finite impact parameter is given by the convolution of the point-source magnification over the source disk. As shown in \citet{1994ApJ...430..505W}, when the source size is much smaller than the Einstein radius ($p\ll1$), the magnification approaches the point-source limit, $\mu \simeq 1/b$ (in the high-magnification regime $b\ll1$). We therefore adopt $b_{\rm thr} \simeq 1/\mu_{\rm thr}$ as the threshold impact parameter.

The differential optical depth scales as $d\tau/db \propto 2b$ for a uniform spatial distribution of lenses. Integrating up to a given threshold $b_{\rm thr}$ gives
\begin{equation}
\tau_{\rm eff} = b_{\rm thr}^2 \, \tau,
\end{equation}
where $\tau$ is the total optical depth defined in Equation~(\ref{eq:tau}). The effective Einstein crossing time is shortened to
\begin{equation}
t_{{\rm eff}, E} = b_{\rm thr} \, t_E,
\end{equation}
because the source must pass within a transverse distance $b_{\rm thr}\theta_E D_L$ rather than the full $\theta_E D_L$. Consequently, the effective event rate after accounting for the impact-parameter requirement becomes
\begin{equation}
N_{\rm eff, det} = b_{\rm thr} \, N_{\rm det}.
\end{equation}
With $b_{\rm thr}=0.1$ (corresponding to $\mu_{\rm thr}=10$), the event rate is suppressed by a factor of 0.1 relative to the estimate in Equation~(\ref{eq:ndet}).

%% For this sample we use BibTeX plus aasjournalv7.bst to generate the

\bibliography{sample701}{}
\bibliographystyle{aasjournalv7}



\end{document}